# Particle Energization in an Expanding Magnetized Relativistic Plasma

Edison Liang

*Rice University, Houston TX 77005-1892*

Kazumi Nishimura, Hui Li and S. Peter Gary

*Los Alamos National Laboratory, Los Alamos, NM 87545*

Using a 2-1/2-dimensional particle-in-cell (PIC) code to simulate the relativistic expansion of a magnetized collisionless plasma into a vacuum, we report a new mechanism in which the magnetic energy is efficiently converted into the directed kinetic energy of a small fraction of surface particles. We study this mechanism for both electron-positron and electron-ion ($m_i/m_e=100$, $m_e$ is the electron rest mass) plasmas. For the electron-positron case the pairs can be accelerated to ultra-relativistic energies. For electron-ion plasmas most of the energy gain goes to the ions.

PACS numbers: 52.65.Rr  52.30.-q  52.65.-y  95.30Qd

An outstanding problem in astrophysics is the acceleration of high-energy particles. The challenge is to find natural mechanisms which can efficiently convert bulk energy, whether it is magnetic, bulk motion, thermal or gravitational energy, into the relativistic energy of a small number of nonthermal particles. Here we report the results of PIC simulations [1], which suggest a new mechanism for the energization of relativistic particles via magnetic expansion.

When a hot, magnetized, collisionless plasma with small gyroradii is suddenly released into a vacuum and the expansion is mainly normal to the magnetic field, the surface gradient generates a strong diamagnetic current, which shields the interior field and confines the field to a thin surface layer. This in turn traps the expanding particles to the surface region. When the expansion is relativistic the surface magnetic field induces a cross electric field with $|\mathbf{E}| \sim |\mathbf{B}|$. The resultant relativistic magnetoacoustic pulse (RMA) pulse, which stays in phase with a fraction of the surface particles, continually accelerate them in the propagation direction to higher and higher energies via its pondermotive force [2-3]. We call this mechanism the diamagnetic relativistic pulse accelerator (DRPA). We have performed PIC simulations for both electron-positron ($m_i=m_e$) and electron-ion ($m_i/m_e=100$) plasmas. In the e+e- case both species are energized equally. But in the electron-ion case most of the energy gain goes to the ions. We have studied both slab and cylindrical geometries with axial fields. While the early behavior of our cylindrical case is qualitatively similar to the slab case, grid-size limitations prevented us from running to sufficiently late times to compare with the slab case. Here we first focus on the slab results.

A potential application of relativistic expanding magnetized plasmas is to gamma-ray bursts [4] and astrophysical jets. Previous works on the expansion of collisionless plasmas into ambient magnetic fields [5-8] focus on the nonrelativistic regime and plasma instabilities [7,8]. To our knowledge there has been no prior study of particle energization via relativistic expanding magnetic fields. Other recent simulations of pair plasmas focused on the structure of collisionless shocks [9-10].

In our code [11] we solve the Lorentz equations of motion of plasma particles and Maxwell's equations. We use 2-1/2 dimensional explicit simulation scheme based on the particle-in-cell (PIC) method for time advancing of plasma particles and fields [1]. Spatial grids are introduced to calculate the field quantities, and the grid separations are uniform, $\Delta x = \Delta z = \Lambda_e$, where $\Lambda_e$ is the electron Debye length defined by using the initial condition. We treat plasma particles as superparticles, with 166 superparticles per cell for each component (electrons and positrons, or electrons and ions).

We use a doubly periodic system in x and z directions, with the system length $L_x=240.\Delta x$ and $L_z= 10.\Delta z$, respectively. Initially, the electron and positron (or ion) distributions are assumed to be Maxwellian with spatially uniform temperature, $T_e$, $T_p$ or $T_i$. The spatial distributions of plasmas have a slab form with width = $0.05 L_x$, height = $L_z$, and the plasma slab is located in the center of the grid. The initial background magnetic field $\mathbf{B_0}=(0, B_0, 0)$ exists only inside the plasma. The initial plasma parameters are: mass ratio $m_p/m_e=1$ (electron-positron case) and $m_i/m_e=100$ (electron-ion case); temperature of plasmas $kT_e = kT_p = kT_i = 5$ MeV; $\Omega_{pe}/\Omega_e=0.105$, where the electron plasma frequency $\Omega_{pe}=(ne^2/\varepsilon_0 m_e)^{1/2}$, the electron cyclotron frequency $\Omega_e=eB_0/m_e$, and the plasma $\beta$ is $\beta_e=\beta_p=\beta_i=0.216$ ($\beta=2\mu_0 nT/B_0^2$). The boundary condition with $L_z =100$ $c/\Omega_e$ may affect particle acceleration at $\Omega_e.t >10^4$ when the relativistic electron gyroradii become $>100$ $c/\Omega_e$. But this does not affect the results reported here. Runs with larger $L_z$ grids show the same results so far but these are continuing. We also know of no potentially unstable modes with wavelengths $\geq 100$ $c/\Omega_e$ that may be impacted by this boundary condition.

Fig.1a shows the time evolution of the global magnetic, electric and particle energies. Starting with equal initial magnetic and particle energies we see that the expansion converts ~80% of the magnetic energy into particle energy at the end of our simulation ($t\Omega_e=1000$). The expansion can be divided into three phases. In phase 1 ($t\Omega_e \leq 80$) magnetic energy is converted into electric field energy with little particle energization as the surface diamagnetic currents build up. In phase 2 ($80 \leq t\Omega_e \leq 500$) both magnetic and electric energies decrease and are converted into particle energy. Finally in phase 3 ($t\Omega_e \geq 500$) the field energies decline slowly with corresponding slow increase in particle energy. Even in this late phase the maximum energy of the most energetic particles continues to increase with time (cf. Fig.3a below). But a decreasing fraction of particles is accelerated as the expansion proceeds.

Fig.2 shows snapshots of the magnetic field, electric field, current density and particle density profiles. We see that as the plasma expands, the magnetic field rapidly diminishes in the interior due to the build-up of the surface diamagnetic current. This in turn confines the particles and field into narrow surface layers. A transverse electric field is induced by the relativistically expanding magnetic field with $E_z \sim B_y$. Those surface particles moving in phase with the narrow RMA pulse is continually accelerated by the pondermotive force of this pulse [2-3]. As time goes on a decreasing fraction of the energetic particles remains in phase with this RMA pulse, accounting for the increasingly narrow and sharp spikes of the phase plots at late times (Fig.3a). Fig.4a gives snapshots of the global particle momentum spectra, showing the development of a nonthermal high energy tail. Our runs end at $t\Omega_e = 1000$ due to our grid size, when $<p_{xmax}> \sim 100$ $m_e c$ at the surface (Fig.3a). But we see no reason why $<p_{xmax}>$ cannot go higher if our simulation continues, as long as there are enough surface particles to sustain the pulse.

The electrodynamics of the electron-ion case is more complex, even though the net result is similar: acceleration of surface electrons and ions to high energies but with the ions gaining most of the energy. Fig.1b shows the time histories of the global magnetic, electric and particle energies. At the end of our simulation ($t\Omega_e = 1000$), ~ 70% of the magnetic energy is converted into ion energy. We can divide the expansion into four phases. Phase 1 ($t\Omega_e \leq 80$) is identical to phase 1 of the e+e- case, in which the magnetic energy is transformed into electric field energy with little particle energization. Phase 2 ($80 \leq t\Omega_e \leq 180$) is similar to phase 2 of the e+e- case with both electrons and ions gaining energy at the expense of the magnetic and electric fields. However, since $m_e \ll m_i$ the relativistic electrons outrun the ions, creating charge separation and an $E_x$ which in turn decelerates the electrons and accelerates the ions. This causes the decline of the total electron energy in phase 3 ($180 \leq t\Omega_e \leq 500$). Finally in phase 4 ($t\Omega_e \geq 500$), the total electron energy stays ~ constant, because the drag on the fast electrons by the ions is in rough balance with the surface acceleration by the RMA pulse, while the ions continue to gain energy from the pull of the fast electrons.

Fig.5 shows the snapshots of the magnetic field, electric field, current, electron and ion density profiles. The propagation of the surface pulse is similar to that in the e+e- case, except that the surface particles are now mostly electrons, forming a nonneutral plasma. At $t\Omega_e$ ~300 a second diamagnetic current is formed near the ion front. However no inductive electric field is associated with this second current because the magnetic field there is too weak. We also find no evidence of the lower hybrid drift instability (LHDI) [7,8] at the ion front because the magnetic field there is weak. This is consistent with the prediction of Ref.[7] that LHDI is suppressed when $\beta$ is sufficiently large. We are continuing to investigate the conditions under which LHDI may occur with additional simulations.

The phase plots (Figs.3bc) show that at late times there are two distinct electron and ion populations. There is a fast electron population at the surface that is accelerated by the EM pulse but decelerated by the ion electric field, and a slow electron population comoving with the ions (Figs.3b & 5d). There is also a fast ion population which is accelerated by the pull of the fast electrons, and a slow population of free-streaming ions in the interior (Figs.3c & 5e). The pull of the fast electrons on the ions does not significantly increase the maximum ion energy. Instead it draws more and more slow ions into the fast population (see the spreading of the ion phase distribution in Fig.3c). But the electrons get increasingly bifurcated in space due to the opposing pulls of the ions and the RMA pulse (Fig.3b). In these runs the ions reach a maximum momentum of ~250$m_e$c, or ion energy of 2.7$m_i c^2$. Hence this mechanism can *accelerate initially nonrelativistic ions to relativistic energies*. The fast electrons still reach a maximum <$p_{xmax}$> of ~100$m_e$c, as in the e+e- case. But here they make up a much smaller fraction of the total population. Thus the electrons globally gain little energy. Figs.4bc show the evolution of the global electron and ion momentum distribution. At late times the momentum distribution is highly anisotropic (<$p_x$> >> <$p_y$> > <$p_z$>), since the acceleration is preferentially in the expansion direction.

We have demonstrated that a relativistic collisionless β≤1 plasma expanding normal to its own magnetic field can efficiently convert the magnetic energy into the directed energy of a small number of surface particles, driving them to ultra-relativistic energies. This DRPA mechanism works for both electron-positron and electron-ion plasmas, though in the latter case most of the energy gain goes to the ions. While we present only the slab case with an initial β~1 plasma, we have performed other simulations with cylindrical geometry, different β and temperature, with qualitatively similar results at early times. The DRPA seems a robust mechanism at least for 2-D field geometries with β≤1 and kT≥$m_e c^2$. Comprehensive parameter studies of DRPA with different kT, β, $m_i/m_e$ and geometry will be reported later. Though our simulations allow seed numerical fluctuations in the initial conditions, we find no evidence of any 2-D plasma instability, including the lower-hybrid drift instability [7, 8], at least for wavelengths ≤ 100 c/$\Omega_e$, (our grid size in z). Admittedly our simulations are restricted to geometries with aligned gradients. Strong nonaligned gradients may cause other 3-D instabilities and coupling of axial and radial motions, which are not represented here. These and other 3-D issues must await simulations with parallel 3-D PIC codes.

EPL is supported by NASA grant NAG5-9223 and LLNL contract B510243. The LANL portion of this work was carried out under the auspices of the US DOE and supported by the LDRD program.

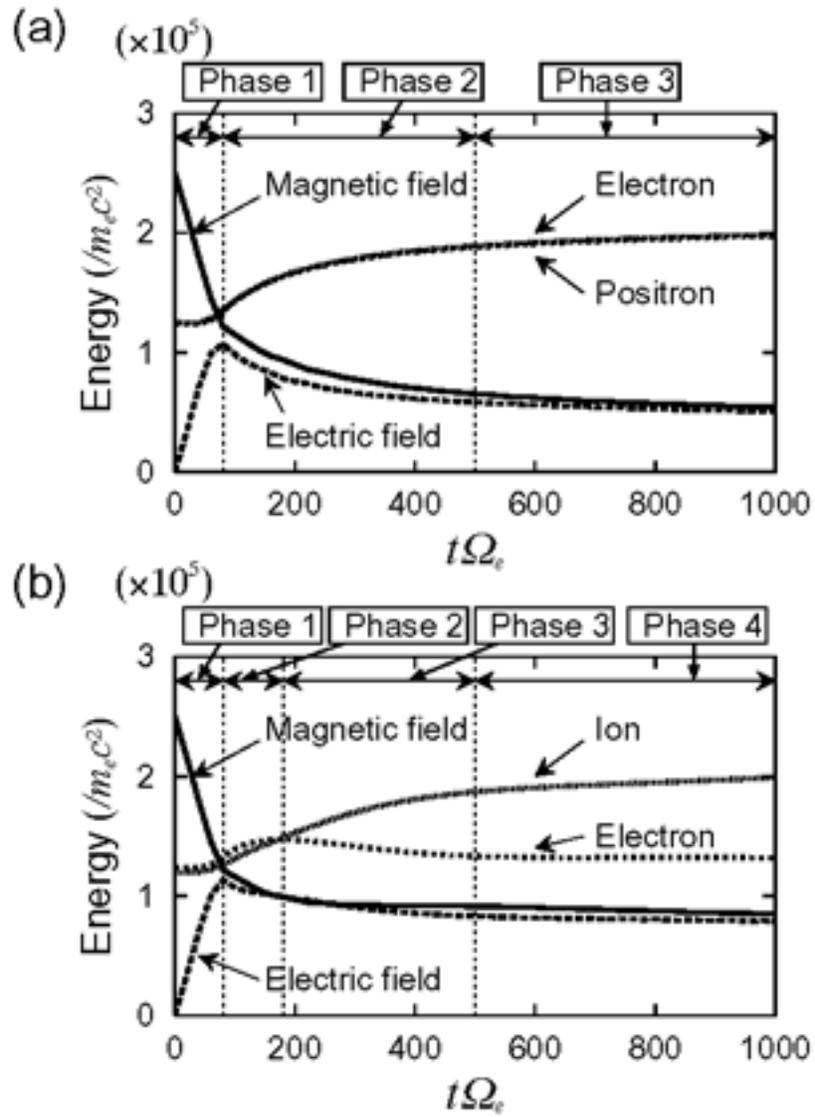

Figure 1 Evolution of the magnetic field, electric field, and particle energies for (a) the electron-positron case; (b) the electron-ion case.

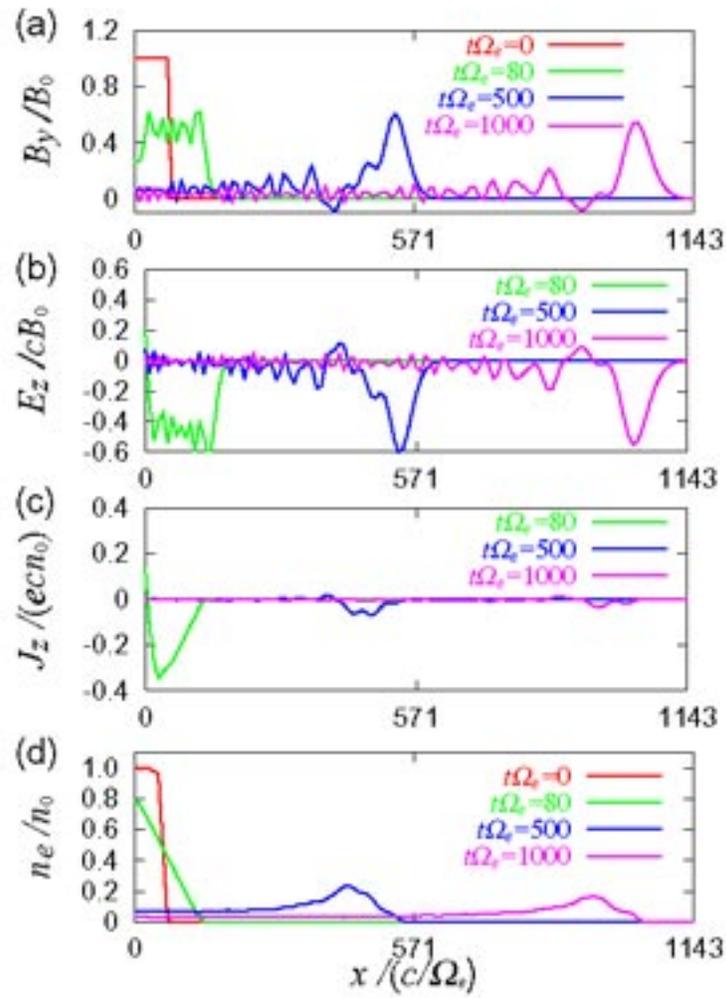

Figure 2 Results for the electron-positron case. Shown are the (a) magnetic field $B_y$, (b) electric field $E_z$, (c) current density $J_z$, and (d) electron density profiles at $t\Omega_e = 0$, 80, 500, and 1000 for $x \geq 0$. Results for $x \leq 0$ are identical. The values of $E_z$ and $J_z$ are zero at $t\Omega_e = 0$.

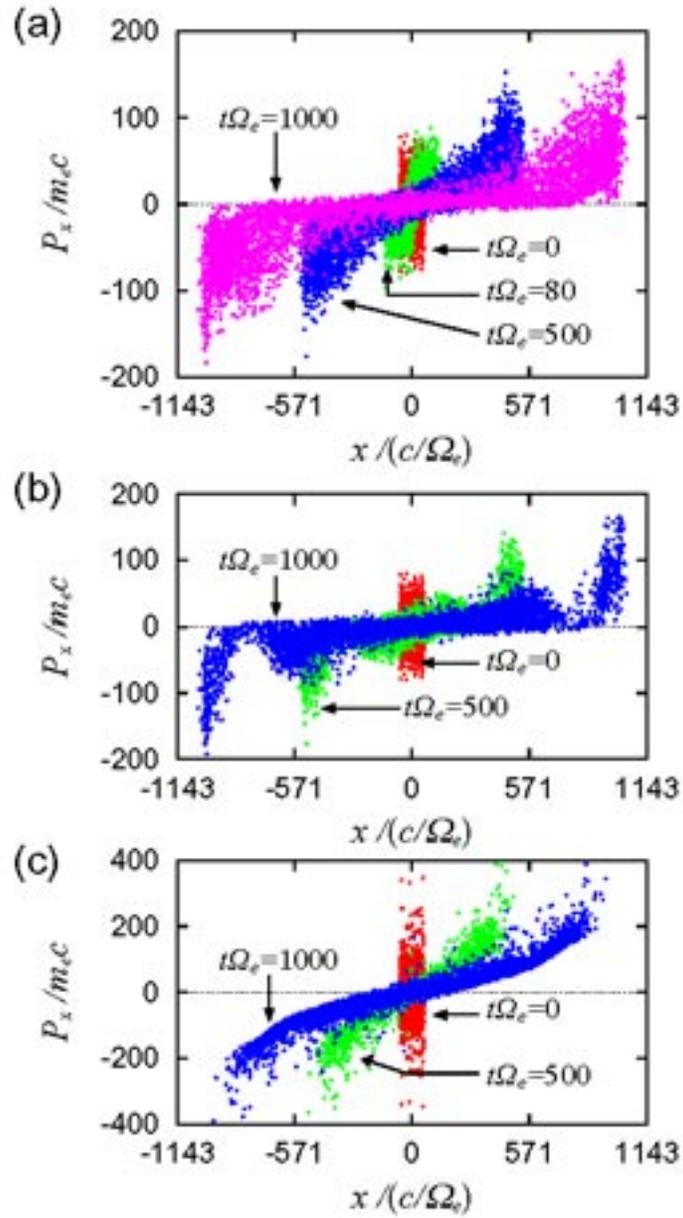

Figure 3 (a) Phase plots for the electron-positron case: we plot the electron distribution in phase space x-$p_x$ at $t\Omega_e$ = 0, 80, 500, and 1000; (b, c) phase plots for the electron-ion case: shown are the electron (b) and ion (c) distributions in phase space x- $p_x$ at $t\Omega_e$ = 0, 500, and 1000.

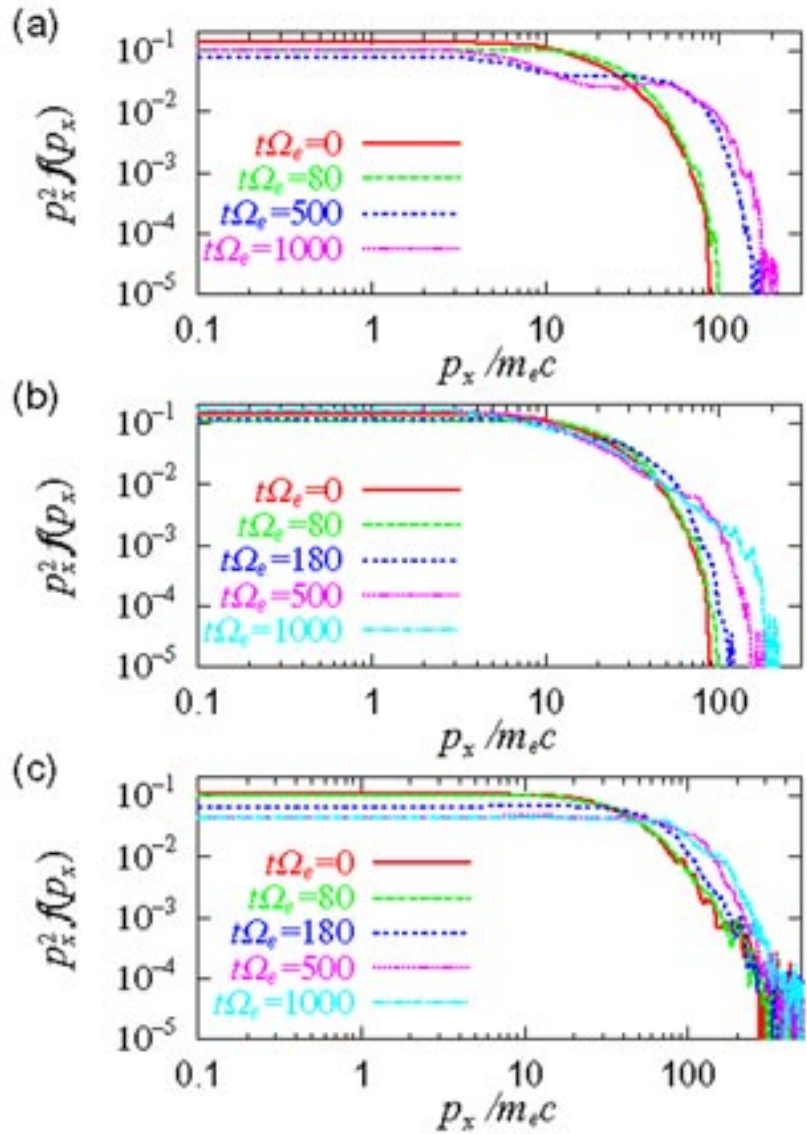

Figure 4 (a) Momentum distributions for the electron-positron case. We plot the electron momentum distributions at $t\Omega_e$ = 0, 80, 500, and 1000; (b,c) momentum distributions for the electron-ion case. We plot the electron (b) and ion (c) momentum distributions at $t\Omega_e$ = 0, 80, 180, 500, and 1000.

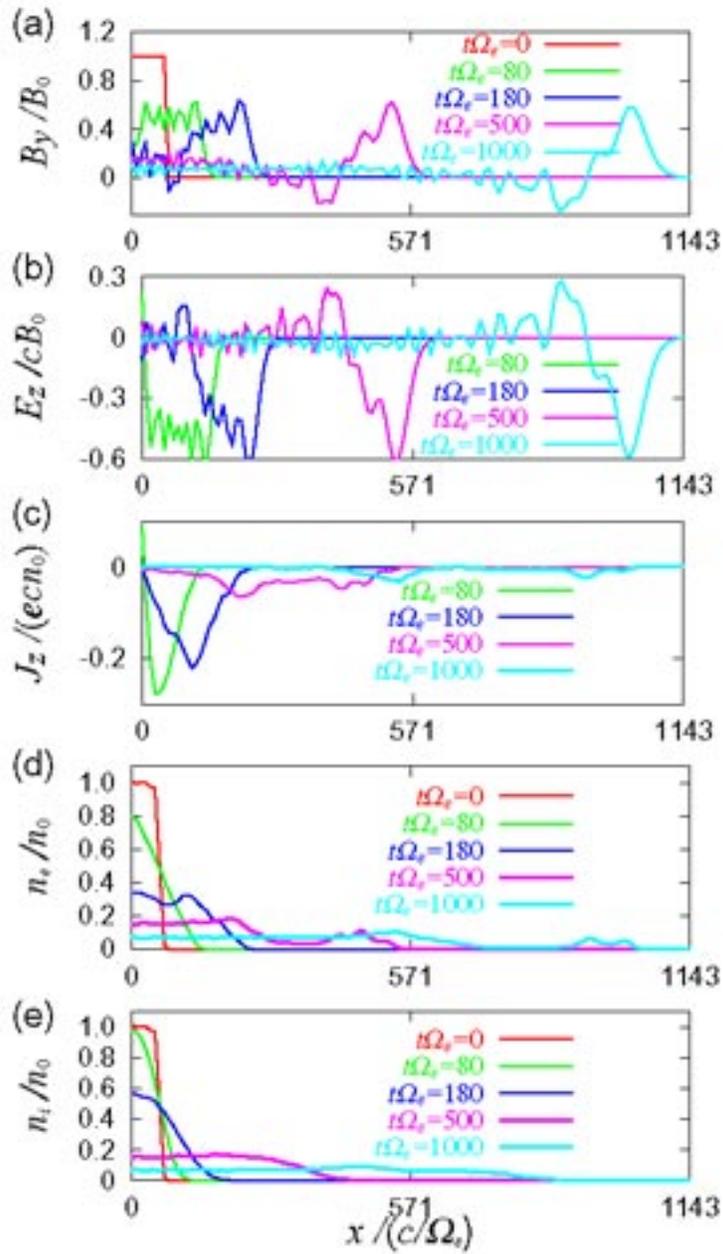

Figure 5 Results for the electron-ion case. Shown are the (a) magnetic field $B_y$,(b) electric field $E_z$,(c) current density $J_z$,(d) electron density $n_e$, and (e) ion density $n_i$ profiles at $t\Omega_e$ = 0, 80, 180, 500, and 1000 for x ≥ 0. The values of $E_z$, and $J_z$, are zero at $t\Omega_e$ = 0.